\documentclass[%
aip,
jcp,%
 amsmath,amssymb,%
preprint,%
superscriptaddress,
]{revtex4-1}
\setcitestyle{super}
\usepackage{float}
\usepackage{geometry}
\usepackage{setspace}
\usepackage{dcolumn}
\usepackage{graphicx}
\usepackage{bm}
\usepackage[utf8]{inputenc}
\usepackage[T1]{fontenc}
\usepackage{mathptmx}
\usepackage{color}
\newcommand{\PE}{\textit{n-}C\textsubscript{720}H\textsubscript{1442}~}
\newcommand{\eg}{\textit{e.g.,}~}
\newcommand{\ie}{\textit{i.e.,}~}
\begin{document}
\title{Divining the Shape of Nascent Polymer Crystal Nuclei} 

\author{Kyle Wm. Hall}
\email{k.wm.hall@temple.edu}
\affiliation{Department of Chemistry, Temple University, Philadelphia, Pennsylvania 19122, United States}
\affiliation{Institute for Computational Molecular Science, Temple University, Philadelphia, Pennsylvania 19122, United States}
\affiliation{Department of Materials Chemistry, Nagoya University, Furo-cho, Chikusa-ku, Nagoya 464-8603, Japan}
\author{Timothy W. Sirk} 
\affiliation{U.S. Army Research Laboratory, Aberdeen Proving Ground, Maryland 21005, United States}
\author{Simona Percec}
\affiliation{Department of Chemistry, Temple University, Philadelphia, Pennsylvania 19122, United States}
\author{Michael L. Klein} 
\affiliation{Department of Chemistry, Temple University, Philadelphia, Pennsylvania 19122, United States}
\affiliation{Institute for Computational Molecular Science, Temple University, Philadelphia, Pennsylvania 19122, United States}
\author{Wataru Shinoda}
\affiliation{Department of Materials Chemistry, Nagoya University, Furo-cho, Chikusa-ku, Nagoya 464-8603, Japan}

\begin{abstract} 
We demonstrate that nascent polymer crystals (\ie nuclei) are anisotropic entities, with neither spherical nor cylindrical geometry, in contrast to previous assumptions. In fact, cylindrical, spherical, and other high symmetry geometries are thermodynamically unfavorable. Moreover, post-critical transitions are necessary to achieve the lamellae that ultimately arise during the crystallization of semicrystalline polymers. We also highlight how inaccurate treatments of polymer nucleation can lead to substantial errors (\eg orders of magnitude discrepancies in predicted nucleation rates). These insights are based on quantitative analysis of over four million crystal clusters from the crystallization of prototypical entangled polyethylene melts. New comprehensive bottom-up models are needed to capture polymer nucleation.  
\end{abstract}

\maketitle

\section*{Introduction}
\begin{figure}[t!]
\includegraphics[width=\textwidth]{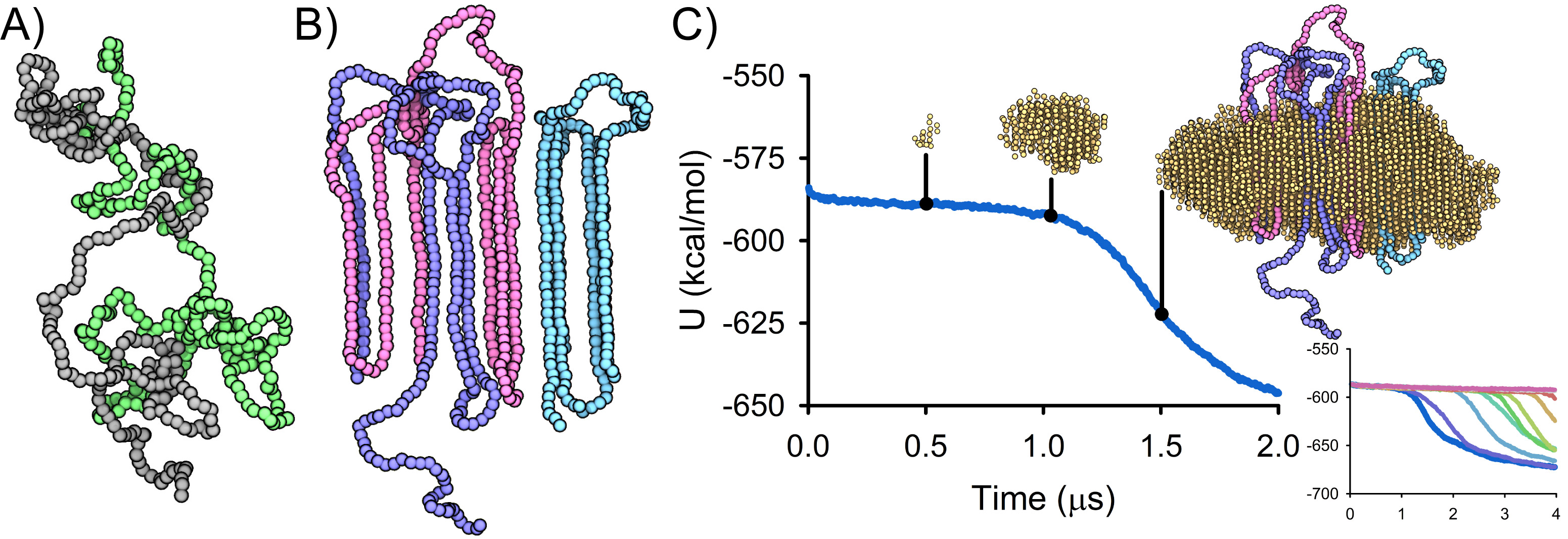}
\caption{Polyethylene chain conformations and early-stage crystallization. A) Two coiled chain conformations from a high-temperature melt prior to crystallization. The chains are from the high-temperature melt that was used to generate the crystallization simulations for this study. B) Three folded chains from a nascent lamella during crystallization in an entangled polyethylene melt, as revealed by one of the simulations from this study. C) The time evolution of the average molecular potential energy (U) as crystallization proceeds. Polymer crystallization coincides with a substantial decrease in potential energy as demonstrated by the adjoining clusters, which are visualizations of the largest crystalline polymer clusters in the system at the specified times. Crystalline coarse-grain beads correspond to the yellow spheres. The purple, pink, and blue chains in the nascent lamella to the right are the same chains as shown in Panel A. The inset to the right shows the average potential energy curves for the ten crystallization simulations for this study; the blue curve is the same simulation in both graphs. Crystallization did not occur in all of the simulations during the $\sim$4-$\mu$s simulation window as evidenced by some of the potential energy curves lacking a pronounced drop.}
\label{fig:PECurves}
\end{figure}

Understanding and controlling the crystallization of semicrystalline polymers, such polyethylene, are of great economic significance given the integral role of crystallization in polymer processing, and the large-scale production of semicrystalline polymers.\cite{Geyer2017}\textsuperscript{,}\footnote{Ref.~\citenum{Geyer2017}. presents compiled data on the production and utilization of plastics. Important plastics that correspond to semicrystalline polymers include polyethylene, polyethylene terepthalate, polypropylene, and many polyamides. Not all plastics are necessarily semicrystalline polymers} Since Keller's seminal work\cite{KellerPhilMag1957,KellerDFS1958} demonstrating that polyethylene crystallizes by adopting chain-folded structures (see Fig.~\ref{fig:PECurves}A-B) such that well-aligned crystalline chain segments called stems form lamellae (Fig.~\ref{fig:PECurves}C), numerous studies\cite{YamamotoMacro2019,HallJCP2019,Kiran2018,CuiChemRev2018,Xiao2017,LuoMacro2016,Kimata1014,LuoSommer2014,YiEtAl_Macromolecules_2013,Muthukumar2000,DoyeFrenkelJCP1999,DoyeFrenkelJCP1998,Liu1998,Hoffmann1997,Sadler1987,SadlerGilmer1986,Barham1985,Lauritzen1960} have explored the crystallization of semicrystalline polymers. 
Crystal nucleation corresponds to the very earliest stages of crystal formation during which a nascent crystal phase (\ie a nucleus) arises from fluctuations in a metastable melt or solution. Despite decades of work on polymer crystal nucleation, there remain many unanswered questions.\cite{CuiChemRev2018} In this paper, we explore the structural evolution of polymer crystal nuclei, especially nucleus shape.

As illustrated in a recent review,\cite{CuiChemRev2018} experimental work with disparate techniques has revealed the multi-faceted, non-classical nature of polymer crystal nucleation under flow conditions; however, experimental work has generally probed the details of whole polymer samples rather than those of individual nuclei. Moreover, experimental descriptions of nucleus and precursor shapes remain relatively qualitative in nature (\eg point-like and thread-like for mild and strong flow conditions, respectively). A detailed description of nucleus shape is needed; many shapes might be seemingly ``point-like'' depending on the scale of observation. In fact, some studies\cite{OkadaHikosaka2013,OkadaPolymer2007} have claimed that polymer nuclei are capable of adopting ``all possible shapes'' based on results from synchrotron small angle X-ray scattering (SAXS). Computational and theoretical studies similarly provide an unclear perspective on the structure of nascent polymer crystal clusters. Some previous studies on polymer crystal nucleation\cite{YiEtAl_Macromolecules_2013,MuthukumarACP2004,MuthukumarPhilTrans2003} and crystal growth\cite{KundagramiJCP2007} have explicitly assumed that nascent polymer crystal clusters exhibit cylindrical geometry while other work implicitly presupposed shape characteristics of polymer nuclei by conjecturing that small clusters exhibit anisotropic interfacial free energies.\cite{Lauritzen1960}  In contrast, qualitative results from previous work\cite{WelchJCP2017} on crystal nucleation in molten polyethylene indicate that nuclei are rough, and there has been acknowledgment that nuclei may adopt irregular shapes.\cite{MuthukumarACP2004} For the crystallization of small molecules, crystal clusters are generally assumed to be spherical (\ie isotropic), and some previous studies have suggested that the same applies to polymer nuclei (\eg see ref.~\citenum{YamamotoMacro2019} and references therein for a discussion of this point). Still other work\cite{SommerPolySci2010,YamamotoJCP2008} has claimed that polymer nuclei are bundle-like aggregates of stems based on qualitative visual inspection of polymer nucleation simulations, and there are indications that the crystallization of an isolated chain proceeds via nuclei corresponding to bundles of stems.\cite{Liu1998} In the context of cold crystallization wherein an amorphous glassy polymer system is heated above its glass transition temperature to induce crystallization, Yamamoto demonstrated that crystallites possess lateral dimensions that are of the same order of magnitude as their lamellar thickness\cite{YamamotoJCP2010} while visual analysis indicated that crystallites are irregularly shaped.\cite{YamamotoJCP2013} However, this work\cite{YamamotoJCP2010,YamamotoJCP2013} did not directly track nuclei and the nucleation process, and it remains unclear how cold crystallization relates to polymer crystallization from melts and solutions.  Many previous \textit{in silico} studies on polymer crystallization\cite{Kiran2018,Xiao2017,LuoPolymer2017,LuoMacro2016,LuoPRL2014,KoyamaEtAl2003,YamamotoJCP2001,WelchPRL2001,Muthukumar2000,YamamotoJCP1998,YamamotoJCP1997} neither considered nor explored the evolution of nucleus shape, focusing instead on other facets of crystallization such as the evolution of polymer chain conformations and/or average stem lengths during crystallization. Still other work has focused on the impacts of flow and extension on polymer crystallization phenomenology.\cite{KoJCP2004,LavinePolymer2003} While there has been much discussion about the structure of polymer nuclei, a direct quantitative assessment of nucleus shape at the molecular level remains outstanding.

Nucleus shape and its evolution have not been assessed directly in part because it remains challenging to study polymer nucleation at the molecular level. Generally, polymer nucleation takes place on short timescales and involves, at least initially, a small number of short polymer chain segments (Fig.~\ref{fig:PECurves}C), thus rendering individual polymer nuclei beyond the reaches of direct experimental investigation. As such, polymer crystallization is being increasingly probed with \textit{in silico} tools, particularly Molecular Dynamics (MD) simulations, which provide fine-grain access in both time and space to the molecular-level details of polymer crystallization (\eg see ref.~\citenum{YamamotoMacro2019,HallJCP2019,AnwarJCP2019,SliozbergMacro2018,Xiao2017,LuoMacro2016,LuoSommer2014, YiEtAl_Macromolecules_2013}). Simulations afford the opportunity to directly and unambiguously probe the molecular-level origins of polymer crystal phases, and their structural evolution during crystal nucleation and growth. Moreover, molecular modeling has previously helped to revise classical thinking on polymer crystallization.\cite{Muthukumar2007} On the basis of extensive MD simulations, this study provides a detailed characterization of polymer nuclei, demonstrating that they are anisotropic entities with neither cylindrical nor spherical geometry, while calling into question additional assumptions.

\section*{Results and Discussion}
MD simulations were leveraged to study polymer crystal nucleation in entangled molten polyethylene, and thus evaluate the evolution of crystal shape during nucleation. These simulations and their subsequent analysis of used standard methodological choices and approaches, so only high-level details of the calculations are presented here with additional information provided in Supporting Information (SI). Polyethylene was represented using the coarse-grain Shinoda-DeVane-Klein (SDK) model.\cite{ShinodaMolSim2007} The SDK model has been previously verified to capture important characteristics of polyethylene (\eg its entanglement mass),\cite{ModelPaper} and it has been previously used to study polyethylene crystallization.\cite{HallJCP2019} The SDK model represents each group of three methylene units along a polyethylene backbone with a single effective coarse-grain bead. All visualizations in this study show SDK coarse-grain beads while all clusters sizes are reported in terms of corresponding numbers of carbon atoms. 
A polyethylene melt comprising \PE chains was equilibrated at 500 K , above the expected melting point of $\sim$400 K for SDK chains.\cite{ModelPaper} Note that chain length considered in this study is within the range of chain lengths of industrial polyethylene. Ten configurations from the high-temperature, entangled melt were then sequentially quenched resulting in ten independent metastable polyethylene melts at crystal forming conditions, specifically 285 K, a temperature that is comparable to what has been considered for polyethylene injection molding.\cite{Leyva2013} Each melt was simulated for $\sim$4-$\mu$s at 285 K and 1 atm.  Additional methodological details are provided in SI 
Subsection A. Not all of the melts crystallized during the $\sim$4-$\mu$s window (see the inset to Fig.~\ref{fig:PECurves}C), indicating that crystallization is initiated by an activated, stochastic process (\ie crystal nucleation) for the conditions considered in this study.

As detailed in SI (Subsection B), polymer nuclei were extracted from the 285 K simulations by assessing the local alignment between polymer chain segments, and then performing cluster analysis on the highly aligned (crystalline) segments. Even though not all of the 285 K melts crystallized during the $\sim$4-$\mu$s simulations, over 4 million crystalline clusters were extracted from saved simulation configurations, including both pre-critical and post-critical clusters. The critical nucleus size corresponds to the location of the free energy barrier to crystallization when using cluster size as the crystallization reaction coordinate as in classical nucleation theory. The critical nucleus size corresponds to $\sim$600 carbon atoms for the polyethylene melts and conditions considered in this study (see SI Subsection C). Previous work\cite{YiEtAl_Macromolecules_2013} has shown that critical nucleus size is independent of chain length during polyethylene crystallization, and so the insights from this study are expected to be relevant to crystallization in polymer specimens composed of much longer chains than those considered in this study (\eg industrial samples). The remainder of this study focuses on clusters in the vicinity of the critical nucleus size (0 - 1200 carbon atoms) in order to establish their structural evolution during the crystal nucleation process.

\begin{figure}[t!]
\centering
\includegraphics[width=\textwidth]{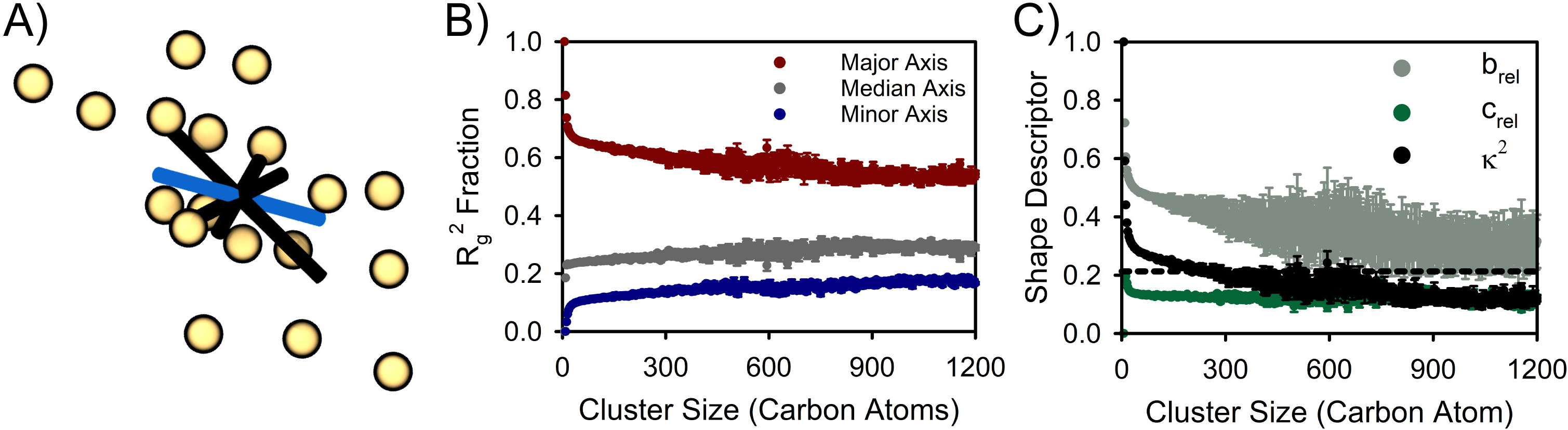}
\caption{Shape characteristics of clusters during polymer nucleation. A) A nucleus extracted from one of the simulations. Yellow spheres indicate constituent coarse-grain beads (chain segments). The principal axes of the cluster correspond to the black lines. The lines are two-fold the semi-axis lengths. The blue line indicates the direction of the nematic director for the constituent chain segments (see SI Subsection F for more details). B) Average fraction of $R_g^2$ accounted for by the squared semi-axis length associated with the major, median and minor axes of nuclei as a function of cluster size. C) Evolution of average $b_{rel}$, $c_{rel}$, and $\kappa^2$ values with cluster size. The dashed black line indicates the average $\kappa^2$ value of clusters corresponding to 30,000-36,000 carbon atoms. Shapes with tetrahedral symmetry or higher symmetry (\eg spheres) have $b_{rel}=c_{rel}=\kappa^2=0$.\cite{TheodorouMacro1985} A prolate cylinder exhibits $c_{rel}=0$ with $b_{rel}>0$ and $\kappa^2>0$. For a perfectly aligned rod of particles, $\kappa^2=1$.\cite{TheodorouMacro1985} The error bars in Panels B and C indicate standard errors.}
\label{fig:ShapeFig}
\end{figure}

Cluster shape was quantitatively assessed using the radius of gyration tensors for the crystalline clusters extracted from the simulations.  More specifically and as detailed in SI (Subsection D), a principal axis system was established for each crystalline cluster using the three eigenvectors of its radius of gyration tensor (see Fig ~\ref{fig:ShapeFig}A); the corresponding eigenvalues are the squared semi-axis lengths of the principal axes. The eigenvalues sum to the squared radius of gyration ($R_g^2$).\cite{TheodorouMacro1985} In this study, the principal axes associated with the largest, smallest, and median eigenvalues are referred to as the major, minor and median axes, respectively. Spherical geometry would require all three axes to contribute equally to $R_g^2$ while cylindrical geometry would require two axes to have the same contribution to $R_g^2$. Neither scenario is observed in Fig.~\ref{fig:ShapeFig}B except for exceedingly small clusters; a cluster composed of only two particles exhibits cylindrical geometry by definition. Rather, the contributions of the major, median and minor principal axes to $R_g^2$ are distinct from smaller clusters to beyond the critical nucleus size. On average, polymer nuclei are neither spheres nor cylinders, but rather anisotropic blobs. 

Shape metrics --- specifically, relative asphericity ($b_{rel}$), relative acylindricity ($c_{rel}$), and relative shape anisotropy ($\kappa^2$) ---  also indicate that small clusters are  neither spheres nor cylinders (see Fig.~\ref{fig:ShapeFig}C and SI Subsection D for the corresponding mathematical details of the shape metrics). Though lower sampling leads to greater noise in the vicinity of the critical nucleus size, there are no apparent transitions in shape metric behaviors. Rather, relative acylindricity values remain comparatively constant while relative shape anisotorpy and relative asphericity values decrease as cluster size increases (Fig.~\ref{fig:ShapeFig}C), implying that larger nuclei are more symmetric. However, polyethylene forms crystalline lamellae, so the trend of decreasing anisotropy with increasing cluster size cannot continue indefinitely. In fact, very large clusters extracted from the simulations exhibit increased anisotropy as indicated by the dashed line in Fig.~\ref{fig:ShapeFig}C. Therefore, polymer clusters exhibit decreasing anistropy as nucleation proceeds, though they are neither spheres nor cylinders, and then undergo a post-critical anisotropy minimum (symmetry maximum) before transitioning to the mesoscopic lamellae characteristic of polyethylene crystals. 

\begin{figure}[t!]
\centering
\includegraphics[width=\textwidth]{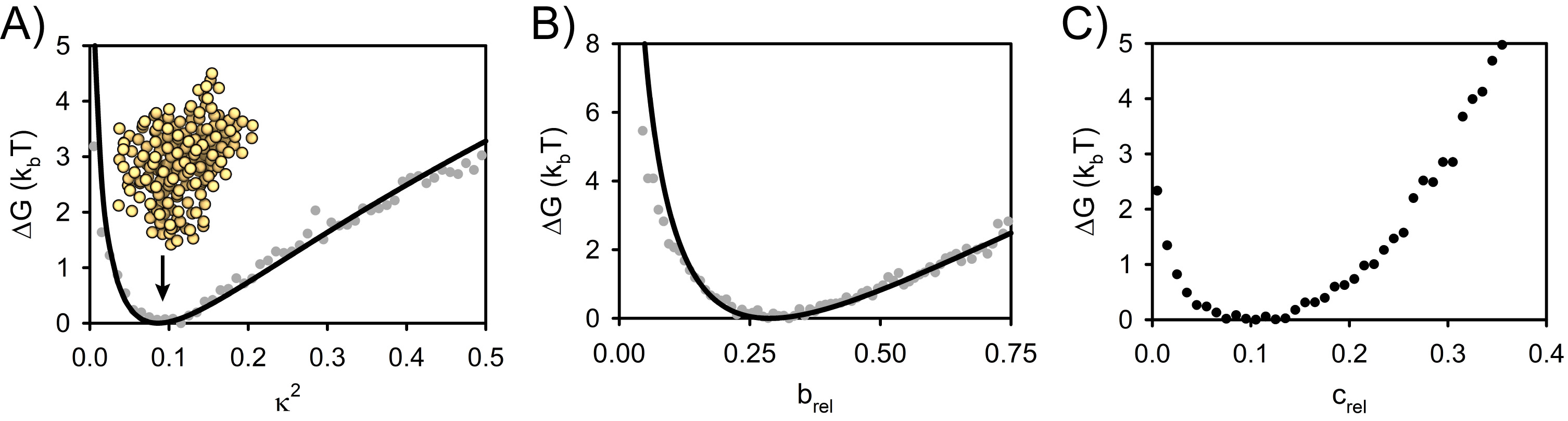}
\caption{Relative free energy profiles in the vicinity of the critical nucleus along: A) $\kappa^2$, B) $b_{rel}$, and C) $c_{rel}$. The cluster in Panel A has shape metric values close to the free energy minima of the profiles.
Each $\Delta G$ profile was constructed by leveraging $G=-k_{b}TlnP(x)$ (where $P(x)$ is the corresponding probability distribution), and then subtracting off the free energy minimum. In Panels A and B, the black lines are the $\Delta G$ profiles based on log-normal fits ($R^2$>0.98) of the underlying probability distributions while the gray points are $\Delta G$ values based on direct conversion of the probability distributions. In Panel C, the black points correspond to the $\Delta G$ profile resulting from direct conversion of the $c_{rel}$ probability distribution; the tails of the $c_{rel}$ probability distribution were not well captured by either log-normal and normal distributions. The underlying probability distributions were constructed using a bin width of 0.01, and nuclei ranging from 300 to 900 carbon atoms in size.}
\label{fig:energyProfiles}
\end{figure}

In order to explore the connections between nucleus shape and the nucleation free energy landscape, relative free energy ($\Delta G$) profiles along $\kappa^2$, $b_{rel}$, and $c_{rel}$ were calculated in the vicinity of the critical nucleus. The free energy profile with respect to $\kappa^2$ is asymmetric about its minimum at $\kappa^2\approx 0.09$ such that it is highly unfavorable ($\Delta G>3k_{b}T$) to form nuclei with lower anisotropy (see Fig.~\ref{fig:energyProfiles}A). Furthermore, it is particularly unfavorable to form spherical nuclei ($b_{rel}$ = 0) as can be seen in Fig.~\ref{fig:energyProfiles}B. Cylindrical geometry, $c_{rel}=0$, is also unfavorable (Fig.~\ref{fig:energyProfiles}C). All three $\Delta G$ profiles coincide with unimodal probability distributions. As such, cylindrical, spherical and higher symmetry nuclei do not correspond to metastable states. There is a preference for anisotropic nuclei. These quantitative findings differ markedly from the shape assumptions made in prior studies,\cite{YiEtAl_Macromolecules_2013,KundagramiJCP2007, MuthukumarACP2004,MuthukumarPhilTrans2003,Lauritzen1960}) and previous claims that that polymer nuclei are essentially isotropic, point-like shapes (see ref.~\citenum{YamamotoMacro2019} and reference therein).

\clearpage

Inaccurate appraisals of nucleus shape could lead to substantial errors when studying and estimating nucleation behavior. For example, consider interfacial free energy estimates. The normalized average squared semi-axis lengths in Fig.~\ref{fig:ShapeFig}B provide the aspect ratios of ellipsoids that approximate the average spatial distribution of constituent particles composing the different size clusters. Comparison of these ellipsoids to spheres of equal volume reveals potential surface area discrepancies of $\sim$6\% for clusters close to the critical nucleus size ($\sim$600 carbon atoms) and over 10\% for smaller pre-critical clusters. A $\sim$6-10\% underestimation in nucleus surface area, arising from treating an anisotropic nuclei as spheres, translates to a potential $\sim$6-10\% overestimation in crystal-liquid interfacial free energies ($\gamma$). Under the basic assumptions of classical nucleation theory, the free energy barrier to nucleation ($\Delta G^*$) is directly proportional to $\gamma^3$, so a $\sim$6-10\% error in $\gamma$ could result in subsequent $\Delta G^*$ calculations being shifted by $\sim$20-33\%. Such $\Delta G^*$ discrepancies could alter estimated nucleation rates by several orders of magnitude as noted by previous work.\cite{HajiAkbari2015}  Moreover, the aforementioned discrepancies are conservative given that the crystal nuclei observed in this study are only approximately ellipsoidal, being rough at the molecular level as discussed later in this study. 

Precise assessments of nucleus structure are required to achieve accurate bottom-up characterizations of nucleation kinetics and thermodynamics, as well as their origins. In practice, nucleus anisotropy may couple to polymer processing factors such as flow. It is also worth noting that there has been recent interest in studying nucleating agents at the molecular level,\cite{BourqueEuroPolymer2018,BourqueJPCB2017} and such agents likely affect structural characteristics of polymer nuclei including shape. Importantly, previous experimental\cite{ZhouNature2019, GasserScience2001} and theoretical\cite{MoroniPRL2005,TruduPRL2006,WangPRE2007} studies on crystal nucleation in metals, colloids and related model systems have observed anisotropic nuclei, suggesting that nucleus anisotropy may occur across disparate crystallization processes. 

The crystal nuclei observed in this study are rough at the molecular level as evidenced by direct inspection of the clusters (see Fig.~\ref{fig:PECurves}C and ~\ref{fig:ShapeFig}A). Previous computational\cite{WelchJCP2017} and experimental\cite{OkadaHikosaka2013,OkadaPolymer2007} studies have also claimed that polymer nuclei exhibit rough interfaces. The roughness and compactness of clusters can be quantified using fractal dimension measurements. The computational details of how fractal dimensions were estimated in this study are provided in SI Subsection E. For reference, a compact volume in 3D possesses a fractal dimension ($D_{f}$) of 3,\cite{Schroeder1991} and values less than 3 correspond to non-compact structures. Note that most clusters of particles at finite temperature are anticipated to display some deviation from $D_{f}=3$ due to thermal fluctuations. For example, interfacial thermal fluctuations (\eg capillary waves) would result in some surface roughness and instantaneous deviations in the surface of an ostensibly spherical droplet in accordance with its interfacial free energy, leading to $D_{f}<3$. However, such fluctuations in a spherical droplet are expected to be small compared to surface area discrepancies noted above. For example, deforming a sphere to increase its surface area by $\sim$6-10\% would require $9k_{b}T$ of work based on previous interfacial free energy estimates for polyethylene (\eg 21.4 mJ/m\textsuperscript{2} from ref.~\citenum{YiEtAl_Macromolecules_2013}) and the average volume of clusters corresponding to 600 carbon atom, the approximate critical nucleus size. Clusters corresponding to 600 carbon atoms exhibit $D_{f}=2.60 \pm 0.13$ (average $\pm$ standard deviation), indicating that nuclei are rough with fractal character.  The rough, fractal nature of early-stage polyethylene crystallites is supported by previous experimental work,\cite{WangPolymer2006} which obtained a small-angle neutron scattering power exponent of -2.2 (\ie $D_{f}=2.2$)  for the early stages of polyethylene crystallization from solution. The difference in the extracted fractal dimensions may be the result of differences in crystallization conditions; this study explores crystallization in entangled polymer melts at 285 K whereas the experimental results\cite{WangPolymer2006} are for short-chain unentangled polyethylene crystallizing from solution at 363 K. Flow may also alter the fractal dimensions of nuclei given that previous \textit{in silico} work\cite{NicholsonJCP2016} on the flow-induced crystallization of C\textsubscript{20}H\textsubscript{42} observed non-compact nuclei with $D_{f}\approx 1.6-2.1$. It is worth noting that the average fractal dimension obtained in this study is comparable to that of proteins (\eg $D_{f}=2.489 \pm 0.172$ from ref.~\citenum{EnrightLeitnerPRE2005}). This reinforces the similarities between synthetic polymers and biological macromolecules, which have been noted by others (\eg see ref.~\citenum{CuiChemRev2018} and reference therein). Consequently, polymer nuclei are not perfectly compact and do exhibit rough surfaces, though they are not unusually rough compared to other molecular structures and systems.\cite{WangPolymer2006,EnrightLeitnerPRE2005,NicholsonJCP2016}

\begin{figure}[b!]
\centering
\includegraphics[width=0.5\columnwidth]{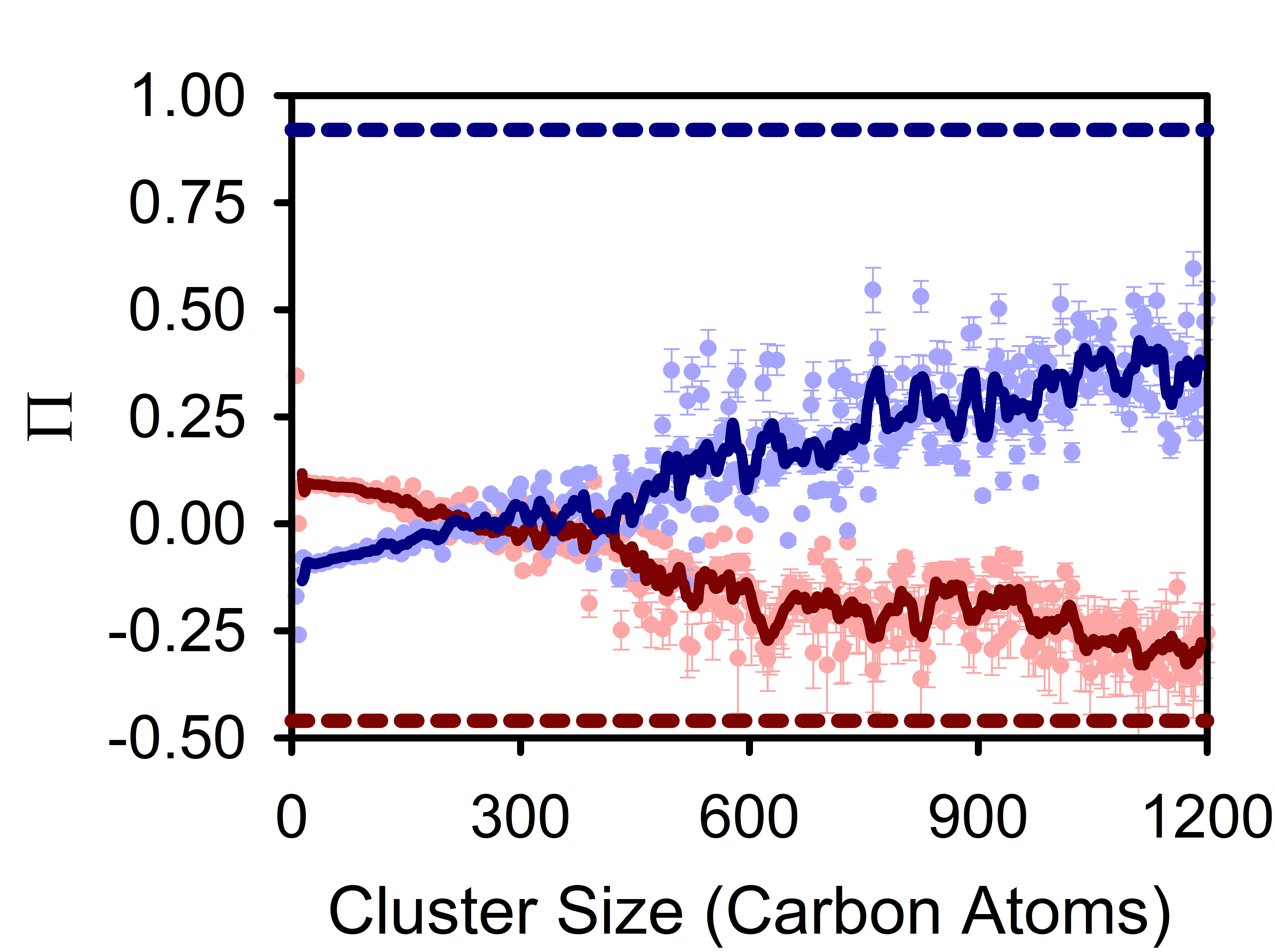}
\caption{The average alignment of the major and minor axes of nuclei with their constituent stems (red and blue data, respectively). The direction of the constituent stems was assessed by determining the nematic director\cite{EppengaFrenkel_MolPhys_1984} of each cluster (see SI Subsection F for details). Average axis-director alignment was then calculated according to $\Pi(n)=<(3cos^{2}\phi-1)/2>_n$ where $\phi$ is the angle between the chosen axis of a cluster and its nematic director, and the angular brackets indicate a conditional average over clusters of size $n$. $\Pi$ values of 1, 0 and -0.5 correspond to perfect alignment, random orientations, and perpendicular orientations, respectively. The dark solid lines are running averages to guide the eyes. The points correspond to underlying data for individual cluster sizes. The error bars indicate standard errors. The dashed lines indicate the average $\Pi$ values for large clusters 30,000-36,000 carbon atoms in size.}
\label{fig:nematicFig}
\end{figure}

For the large-scale lamellae observed in solid-state polyethylene, the thickness of a lamella is small with respect to its lateral extent. Moreover, following crystallization at high-driving force conditions such as those considered in this study, the backbones of the constituent polymer stems are anticipated to be roughly aligned with the lamellar normal.\cite{deSilvaMacro2002,deSilva2003} In turn, if small nuclei are anisotropic blobs structurally consistent with polyethylene lamellae, then the minor axis of a nucleus should be on average approximately parallel to the stems composing the nucleus. Similarly, the major axis should be perpendicular to the direction of the stems. 
Yet, small nuclei do not exhibit this structuring as shown in Fig.~\ref{fig:nematicFig}. For very small nuclei, there is a slight preference for the major axis of the nucleus to align with its constituent chains as indicated by $\Pi>0$ while the opposite is true for the minor axis ($\Pi<0$). Crystallization in small clusters is apparently more likely to proceed initially via propagation along polymer chain backbones rather than through perpendicular inter-segment recruitment. The situation is reversed as the critical nucleus is approached. For nuclei of critical size, there is a preference for the minor axis to be aligned with the constituent chain segments in contrast to the major axis, and these preferences continue to increase as the nuclei grow further (Fig.~\ref{fig:nematicFig}). Large clusters do exhibit the alignment behaviors expected for polyethylene lamellae formed under strong-driving force conditions as revealed by the dashed lines in Fig.~\ref{fig:nematicFig}.

\begin{figure}[tbhp]
\centering
\includegraphics[width=0.5\columnwidth]{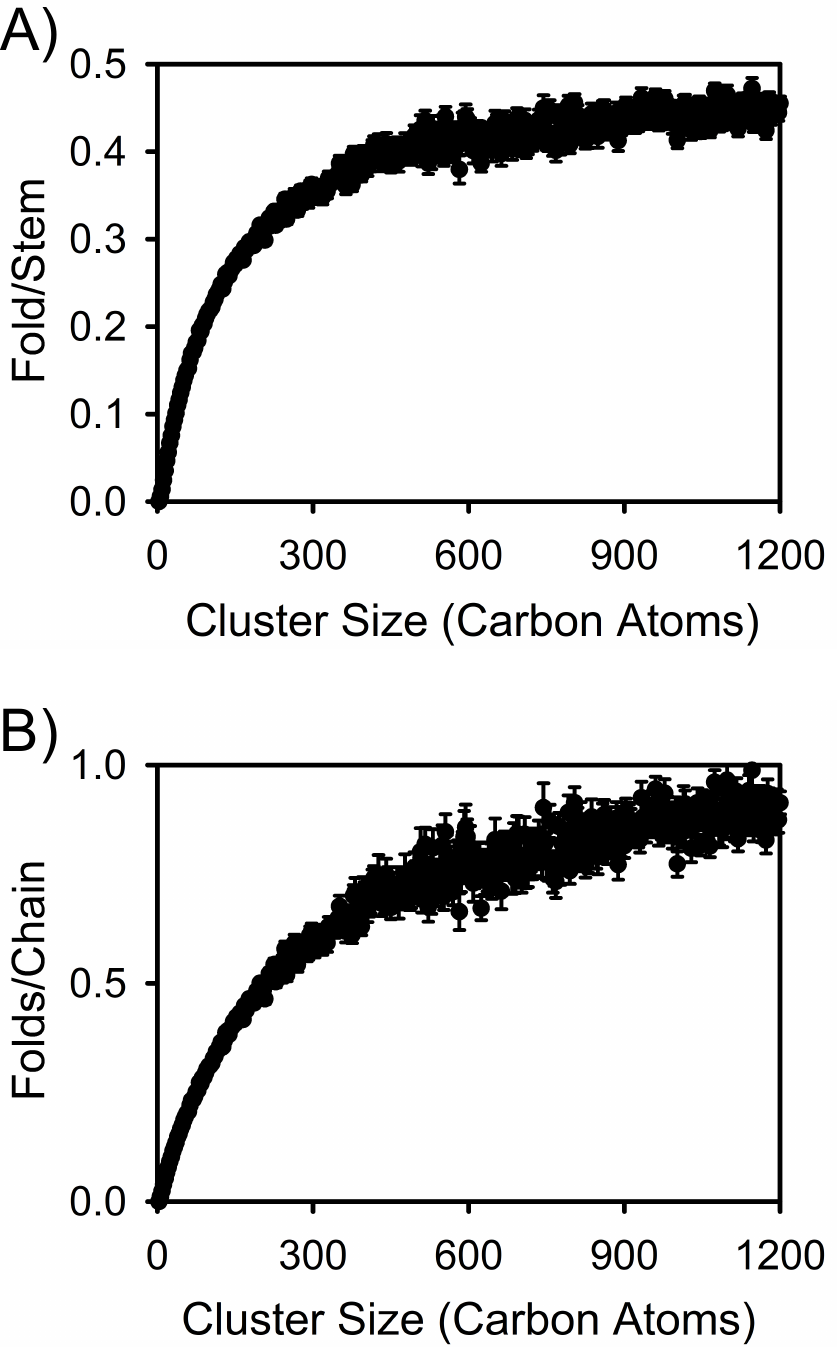}
\caption{Fold properties of polymer nuclei. A) Evolution of average number of folds per stem with cluster size. A stem corresponds to an uninterrupted sequence of crystalline beads belonging to one extracted cluster and one polymer chain. A fold is a non-crystalline series of beads belonging to one chain, and bounded by stems belonging to the same cluster. The maximum value of folds/stem is one for a cluster with all stems terminated by folds. B) Average number of folds per constituent chain as a function of nucleus size. The error bars in both panels correspond to standard errors.}
\label{fig:FoldFig}
\end{figure}

Given that the shape and alignment characteristics of polymer nuclei do not conform to those of polymer lamellae, it is reasonable to expect that other attributes of polymer lamellae may not be applicable to polymer nuclei. For example, it is often assumed that polymer nuclei have well-defined fold surfaces and lateral surfaces just like crystal lamellae.\cite{YiEtAl_Macromolecules_2013,Lauritzen1960} In the lamellar case, the fold surfaces coincide with the upper and lower surfaces of the lamella while the lateral sides correspond to exposed stems at the edge of the lamella. However, early-stage nuclei exhibit very few folds with respect to their number of constituent stems as shown in  Fig.~\ref{fig:FoldFig}A, and chains composing early-stage nuclei do not possess many folds as shown in  Fig.~\ref{fig:FoldFig}B. These findings are consistent with previous results\cite{YiEtAl_Macromolecules_2013} indicating that small nuclei have few folds. As such, early-stage crystal nuclei do not have well-defined fold surfaces, in contrast to lamellae. This aligns with previous work suggesting that small nuclei have a fringed micelle structure.\cite{YamamotoJCP2008}  However, fringed micelle treatments often invoke particular structures for the nucleus and particular orientations of stems with respect to nucleus shape, assumptions that are inconsistent with the results obtained in this study. Even at the critical nucleus size, the constituent chains of a nucleus are on average participating in less than one fold, raising the question of how does the fold surface develop on transitioning from crystal nucleation to growth. Consequently, it is not appropriate to take macroscopic quantities and interfacial properties based on crystal growth studies, and project them onto polymer nucleation processes. Distinctions between nuclei and lamellae may help to explain discrepancies in nucleation-related thermodynamic estimates (\eg interfacial free energies) noted by previous work.\cite{YiEtAl_Macromolecules_2013}

\clearpage
\section*{Conclusions}
Small nuclei are anisotropic blobs, neither spheres nor cylinders, that are structurally distinct from the mesoscopic structures that ultimately arise from polymer crystallization. Moreover, it is thermodynamically unfavorable for nuclei to display cylindrical, spherical or other high symmetry structures in the vicinity of the critical nucleus size. Consistent with this, nuclei are rough, and exhibit fractal dimensions comparable to proteins. The development of fold surfaces and structural characteristics of polymer crystal lamellae are post-critical phenomena. Inappropriate assessments of nucleus shape might give rise to substantial errors and potentially orders of magnitude discrepancies in nucleation rates. As such, the anisotropic blob-like nature of nuclei has important implications for studying and predicting nucleation behavior. Polymer crystal nucleation is a multi-faceted process accompanied by changes in alignment and shape. To achieve transformative molecular-level understanding of polymer crystallization, and thus support initiatives to tailor polymer materials through processing optimization, new comprehensive bottom-up models are likely to be needed.

\section*{Acknowledgments}
This work was supported by the US Army Research Laboratory (contract numbers: W911NF-18-9-0269 and W911NF-16-2-0189). This study used the high-performance computing resources at Temple University, and was thus also supported by the National Science Foundation (major research instrumentation grant number: 1625061). This work was completed in part while K. Wm. Hall was an International Research Fellow of the Japan Society for the Promotion of Science (JSPS); K. Wm. Hall thanks JSPS for their support. M. L. Klein thanks H.R.H. Sheikh Saud for his support via a Sheikh Saqr Research Fellowship.

\bibliography{workcited}

\end{document}